\documentclass[a4paper,noarxiv,twocolumn,12pt]{quantumarticle}
\usepackage[utf8]{inputenc}
\usepackage[english]{babel}
\usepackage[T1]{fontenc}
\usepackage{amsmath}
\usepackage{hyperref}
\usepackage{braket}

\usepackage{tikz}
\usepackage{lipsum}

\usepackage{mdframed}
\usepackage{booktabs}
\usepackage{longtable}
\usepackage{graphicx, caption}
\usepackage{amssymb}
\usepackage[margin=0.5in]{geometry}

\begin{document}

\title{Application of a Quantum Search Algorithm to High-Energy Physics Data at the Large Hadron Collider}

	\author{Anthony E. Armenakas}
	\affiliation{Harvard University, Cambridge, MA}
	\email{anthony.emmanuel.armenakas@cern.ch}
	\author{Oliver K. Baker \hspace{-9mm}\thanks{(Corresponding author)}}
	\affiliation{Department of Physics, Yale University, New Haven, CT}
	\email{oliver.baker@yale.edu}

	\maketitle

\begin{abstract}
We demonstrate a novel method of applying a scientific quantum algorithm, Grover’s algorithm (GA), to search for rare events in proton-proton collisions at 13 TeV collision energy using CERN's Large Hadron Collider. The search is of an unsorted database from the ATLAS detector in the form of ATLAS Open Data. As indicated by the Higgs boson decay channel $H\rightarrow ZZ^*\rightarrow 4l$, the detection of four leptons in one event may be used to reconstruct the Higgs boson and, more importantly, evince Higgs boson decay to some new phenomena, such as $H\rightarrow ZZ_d \rightarrow 4l$. In searching the dataset for collisions resulting in the detection of four leptons, the study demonstrates the effectiveness and potential of applying quantum computing to high-energy particle physics. Using a Jupyter Notebook, a classical simulation of GA, and multiple quantum computers, each with several qubits, it is demonstrated that this application makes the proper selection in the unsorted dataset. The implementation of the method on several classical simulators and on several of IBM's quantum computers using the IBM Qiskit Open Source Software exhibits the promising prospects of quantum computing in high-energy physics.
\end{abstract}

\tableofcontents

\section{Introduction}
Quantum computing (QC) uses the quantum mechanical properties of quantum bits (qubits) to complete tasks with far greater efficiency than its classical computing counterpart. Prior to the current study, this fundamentally distinct form of computing had yet to be applied in the search of particle physics databases in this way. With high-energy particle physics being a data-intensive science, a large amount of data is collected at the Large Hadron Collider (LHC) at CERN, where particles are accelerated and collided at high energies. Detectors such as that of the ATLAS (A Toroidal Lhc ApparatuS) collaboration collect data on the resulting shower of particles. With collisions run at higher energies, and as more and more particle decay channels are discovered, the ability to make novel scientific conclusions with this data requires increasingly improved methods of data sorting, pattern recognition, and data analysis. Grover's quantum algorithm \cite{Grover_1996} \cite{Korepin_2005}, a search algorithm designed to run on quantum computers, has been shown to search over a set of data quadratically faster than classical search algorithms. The current study demonstrates the development and application of a quantum search algorithm to complete tasks and answer essential questions in particle physics---specifically, to search for signals of the Higgs boson decay products in events detected at CERN's LHC, permitting "reconstruction" of the Higgs boson.

Event selection in high-energy particle physics remains a significant challenge to modern scientific research. The eagerly anticipated LHC upgrade will enable the acceleration of particles to unprecedented energies and proton-proton 
({\it pp}) collision frequencies. The latest ATLAS Open Data release contains databases of {\it pp} collision events at a highest-ever 13 TeV collision energy. Large databases, containing vast amounts of event data recorded by the ATLAS detector, must be filtered, sorted, and searched. In this current study, we approach the enormity of these databases in a novel manner---by implementing a quantum search. The augmented computational power of quantum computing as compared to its classical counterpart, which stems from quantum superposition, expedites the search; while classical bits have two possible states, 0 and 1, quantum bits, or qubits, may exist in the measurable $\ket{0}$ and $\ket{1}$ states as well as in a quantum superposition of $\ket{0}$ and $\ket{1}$. Grover's algorithm (GA) selects a target quantum state and increases the probability that the system is measured in that state. An application of GA to the search of an HEP database was developed and run on a variety of quantum computers and simulations. The ATLAS Open Data containing events from 13 TeV {\it pp} collision energy was searched. The database contains data on lepton transverse momenta as detected by the ATLAS collaboration's detector. We searched the database for collisions' products, or events, in which four leptons were detected at the appropriate energy and mass window, evincing the presence of the Higgs boson in the post-collision particle showers, as indicated by the Higgs boson's $4l$ decay channel. 

\section{Methods}
GA was run on simulated qubits in the R programming language and in IBM's Qiskit using Python. The algorithm was then run on the hardware of real IBM quantum computers, producing slightly less distinct peaks in target state measurement probability---the product of quantum decoherence. GA was successfully used to search the ATLAS database for events with four leptons. When run on the simulator, the correct state was selected with 100\% probability, and the code yielded the event(s) with four leptons. However, when run on IBM quantum hardware, quantum decoherence reduced the peak in target state probability. We examined ways in which the decoherence was mitigated. 

\subsection{Search Motivation}
The Higgs boson that was discovered at the LHC 
\cite{atlas2012observation} \cite{cms2012observation}, if it is the standard model (SM) boson, will couple to SM particles in a manner that is unlike any other lepton, quark, or gauge boson; its coupling strength is related to the particle's mass. If the nature of its coupling depends on the mass of the state it couples to, it may provide a new means to search for phenomena that are beyond the SM of particle physics. The Higgs boson could provide a portal to a so-called Dark Sector of new particles and interactions, coupling to them in a unique way that cannot be probed with other SM probes. Completely new physical states may therefore be accessible experimentally, via coupling to the Higgs boson, in a way that did not exist previously. 

Theoretical studies, supported by astrophysical and cosmological experimental
data, indicate that these Dark Sector particles can lead to very rare events in
LHC collisions \cite{Marciano:2013t1} \cite{BNL2012dark} \cite{Curtin:2013fra} 
\cite{gopalakrishna2008higgs}. An LHC Dark Sector search consists of detecting a resonance that decays to leptons.  Leptons are a class of structureless particles with spin-$1/2$ that do not exhibit strong interactions. Leptons can be electrons ($e$), muons ($\mu$), tau particles ($\tau$), or neutrinos 
($\nu_e$, $\nu_{\mu}$, $\nu_{\tau}$). 

In the study described here, we chose to 
search for a resonance that decays to electron-positron or positive muon and 
negative muon pairs.    This resonance  that signals a new particle 
would be seen above a broad spectrum of background SM events.
Electron and positron candidates consist of clusters of energy deposited 
in the electromagnetic calorimeter associated with their tracks inside the detectors.  
The clusters matched to tracks are required to satisfy a set of identification criteria 
that require the longitudinal and transverse shower profiles to be consistent
with those expected for electromagnetic showers. These are registered as hits, 
which are then translated in software to electron and positron energies, momenta, 
charges, and position at any given time.
 
Positive and negative muon candidates are formed by matching reconstructed 
detector tracks in one ATLAS spectrometer subsystem with either complete or partial tracks 
reconstructed in a separate ATLAS subsystem. If a complete track is present, the two 
independent momentum measurements are combined. The particles are identified 
as muons if their calorimetric energy deposits are consistent with a minimum ionising particle.

In this current analysis, the detector data has already been formatted to run using classical
algorithms on a classical computer.  This format had to be changed to suit the quantum algorithm
(GA) that is being used for the results shown here.  It is understood that for future analyses 
that will employ quantum search algorithms such as the one described here, it will be much
more efficient to store that data in a quantum format initially.  This formatting should be 
the next step taken after collecting the tracking hits, energy deposits, shower profiles, etc, as
described above.

\begin{widetext}
\begin{center}	
\begin{figure}[h!]  
\centering
\vspace*{-1 mm}
\centerline{\includegraphics[width=1.0\textwidth, height=2.5 in]{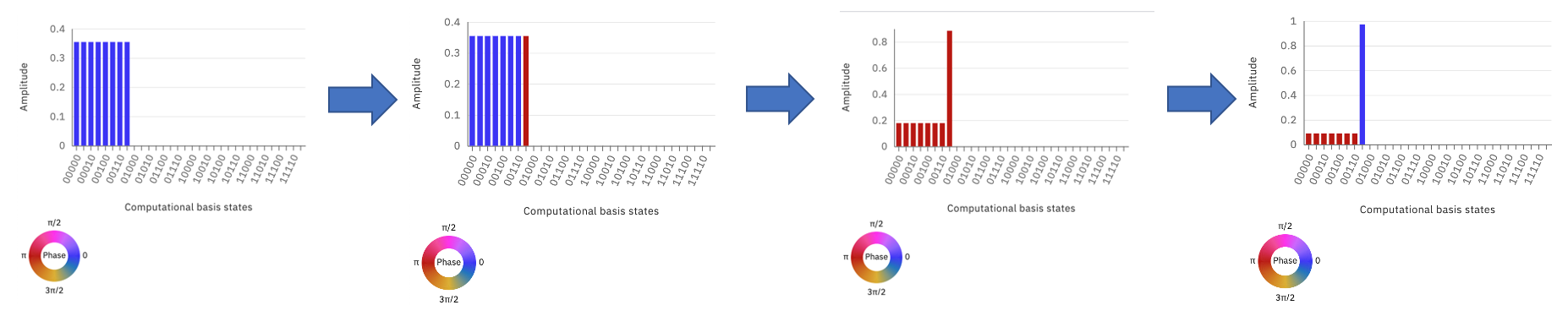}} 
\vspace*{3 mm}
 \caption{{\bf The effect of Grover's algorithm on the the amplitude of each qubit state.} The amplitudes of all 8 possible states of the three-qubit quantum chip are shown at various stages of the algorithm. The system is placed in an equal superposition (far left), with each state's amplitude being $\frac{1}{2\sqrt {2}}$. Progressing to the right, the phase transformation and amplitude amplification of the target state are shown. The process is repeated in a second iteration, resulting in a greater target state amplitude (far right). }
\label{fig:GAp}
\end{figure}
\end{center}
\end{widetext}

The collected information, as described in the
previous two paragraphs, is converted into database entries that are appropriate for the designed 
method of Dark Sector searches using exotic decays of the Higgs boson, and which are in a
format that can be understood and handled by the quantum search algorithm. In this study,
Grover's quantum search algorithm was used, and the database searched contained the transverse 
momenta of detected leptons.  

Data pertaining to the exotic decays of the Higgs boson used in Dark Sector searches was sought.
The Higgs boson may 
decay into a $ZZ^*$ ($Z^0$ boson and an off-shell $Z^0$ boson) pair, which in turn decay into 
four leptons. This decay channel is commonly represented as $H \rightarrow ZZ^* \rightarrow 4l$
\cite{atlas2012observation} \cite{cms2012observation}.
In rare cases, the decay channel $H \rightarrow ZZ_d \rightarrow 4l$, where the $Z_d$ refers
to a Dark Sector vector boson that is beyond the standard model of particle physics, can occur 
\cite{zdark:ATLAS}. 
Once produced, the Higgs boson decays quickly, leaving the four leptons at the end of the decay 
channel to be detected by the ATLAS detector. A search was designed on a database of detected 
leptons; the search was for collisions, or events, in which four leptons were detected. The data 
presented in this existing work was taken from the LHC's recent run at a record-high proton-proton 
collision energy of 13 TeV, where the chances of Higgs boson production are higher.

\subsection{Grover's Algorithm}
Grover's algorithm was a vital part of the search process on real data. Before it was used to develop a 
search of real ATLAS data, Grover's algorithm was run on the qubits of several quantum computing 
simulators and real quantum computers. The algorithm searches a database of N entries, where N is 
some integer. A classical computer (composed of bits, as opposed to qubits),
goes through each entry until the target entry is found. On average, the computer must 
make (N+1)/2 probes before the target entry is selected, assuming an equal probability 
that each entry is the target entry. To demonstrate that on average (N+1)/2 probes are made on a 
classical computer, take a list of N database entries, where each can be the target entry. For target 
entry $i$ in \{1, N\}, the classical computer probes each entry in \{1, $i$\} before making the correct 
selection; thus it takes $i$ probes to reach the correct selection of the target entry $i$. The sum of all 
probes for all potential target entries is $\sum\limits_{i=1}^N i = \frac{N(N+1)}{2}$, and the number of 
potential searches is N, so the average number of probes per search is $\frac{\frac{N(N+1)}{2}}{N} = 
\frac{N+1}{2}$.

\begin{widetext}
\begin{center}
\begin{figure}[h!] 
\centering
\vspace*{-15 mm}
\centerline{\includegraphics[width=0.98\textwidth, height=5. in]{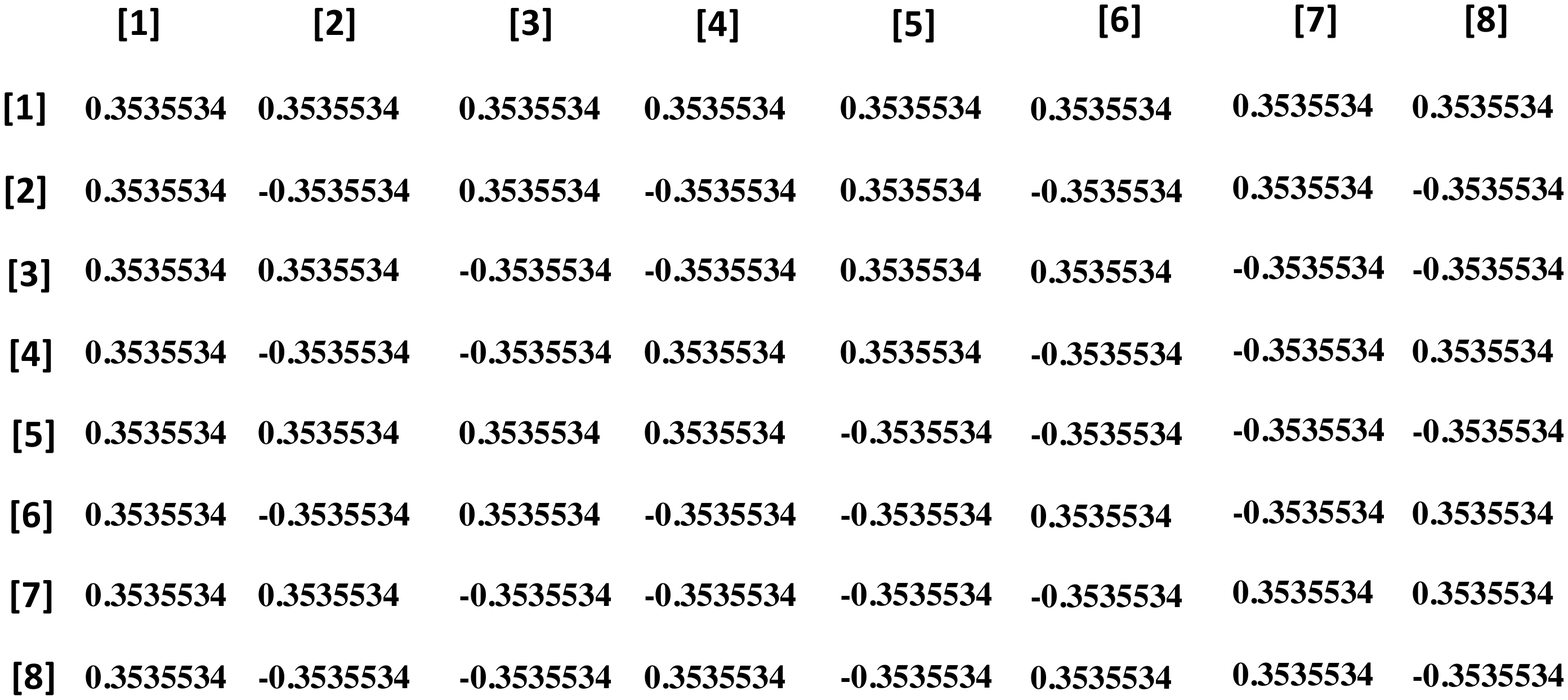}}
\vspace*{-18 mm}
\caption{{\bf QC simulation matrix.} A matrix of qubit wave-function amplitudes in the R Programming QC Simulator.}
\label{fig:QCSM}
\end{figure}
\vspace*{-10 mm}
\end{center}
\end{widetext}

Quantum search algorithms execute the search in 
a fundamentally different manner. On a two-qubit quantum computer, there are four 
measurable states, as each qubit can be measured in the $\ket{0}$ or $\ket{1}$ state (yet may 
exist in a superposition of states prior to measurement). On a three-qubit quantum 
computer, there are 8 measurable states ($\ket{000}$, $\ket{001}$, $\ket{010}$, $\ket{011}$, $\ket{100}$, $\ket{101}$, $\ket{110}$, and $\ket{111}$).
Thus in general, on an n-qubit quantum computer there are 2$^n$ measurable states. In 
quantum superposition, each state has an amplitude, such that a three-qubit state in 
superposition can be written as: 
\begin{widetext}
\begin{equation}
\ket{\Psi} = a\ket{000} + b\ket{001} + c\ket{010} + d\ket{011} + e\ket{100} + f\ket{101} + g\ket{110} + h\ket{111}
\end{equation}
\end{widetext}
where a, b, c, d, e, f, g, and h are the amplitudes of each measurable state in superposition. The 
modulus squared of the amplitude of a state is the probability of its measurement in that state. The sum 
of the probabilities of each state's measurement is 1. Thus, 

\begin{equation} 
{\lvert a \rvert^2} + {\lvert b \rvert^2} +  {\lvert c \rvert^2} + {\lvert d \rvert^2} + {\lvert e \rvert^2} + {\lvert f \rvert^2} + {\lvert g \rvert^2} + {\lvert h \rvert^2 }= 1.
\end{equation}

Grover's algorithm searches for a target state by amplifying the state's amplitude, minimizing the 
amplitudes of all other states. The end result is that the qubit system has a higher probability of being 
measured in the target state \cite{2011:Strubell}.   
The first step of Grover's algorithm is the placement of all states in an equal superposition, where
their amplitudes are equal. Then, Grover's algorithm manipulates the amplitudes of the qubit
system, selecting the target state's amplitude and amplifying it in what is called the "Grover
iteration" \cite{2002:Diao}.  
Figure~\ref{fig:GAp} demonstrates how the Grover iteration alters measurement probability. The three-qubit 
quantum chip is placed in an equal superposition, with each state's amplitude being $\frac{1}{2\sqrt {2}}
$. The target amplitude is multiplied by -1 (a phase shift of $\pi$) before the algorithm performs amplitude 
amplification on the target state. After the second GA iteration, the target state amplitude, and thus the state's 
probability of measurement, is closer to 1. 
 
\subsection{Algorithm Application}
Grover's algorithm was first run in R Programming's quantum computing simulator. 
Figure~\ref{fig:QCSM} shows the matrices with which the QC simulator in R simulates a quantum 
computer. In place of physical changes of qubit amplitudes, the simulator runs mathematical 
transformations.

\vspace*{-0.5 mm}
\begin{figure}[tp!] 
\centering
\includegraphics[width=0.5\textwidth, height=2.8 in]{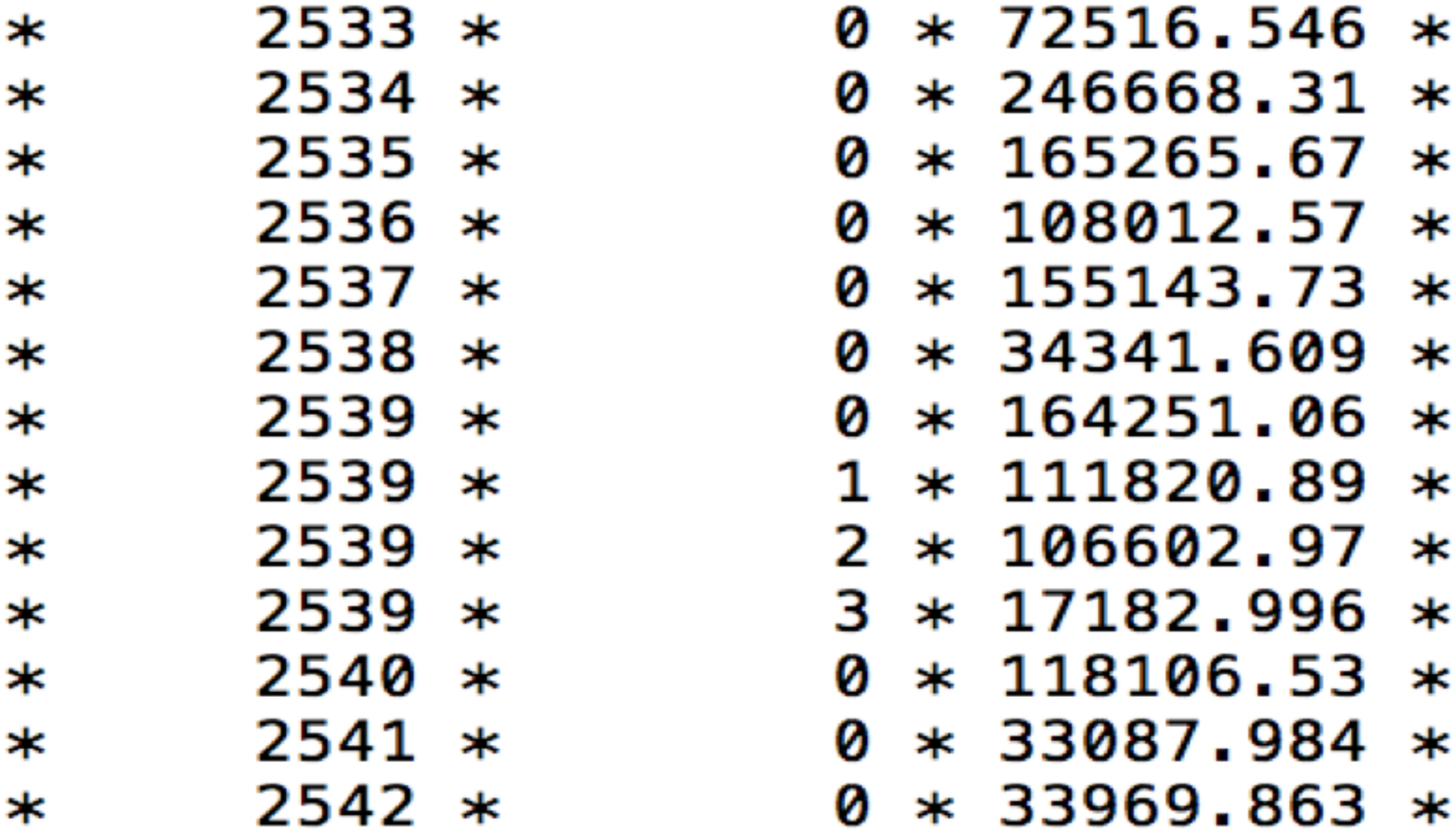}
\vspace*{-10 mm}
\caption{{\bf A sample from the Open Data keyed into the circuit.}}
\label{fig:OpenData}
\end{figure}

Registration with IBM granted access to real quantum computers in addition to QC simulators through 
Qiskit. From then on, a Jupyter Notebook was employed to create and modify code in Python. The 
simulation backend ibmq\_qasm\_simulator was used for testing numerous times before algorithms 
were run on real devices. After running Grover's algorithm in different ways on both real and 
simulated quantum hardware as to develop the best method for application to real encoded data, the 
search of LHC data was developed.   

\begin{figure}[bp!] 
\vspace*{-5 mm}
\includegraphics[width=0.45\textwidth, height=2.8 in]{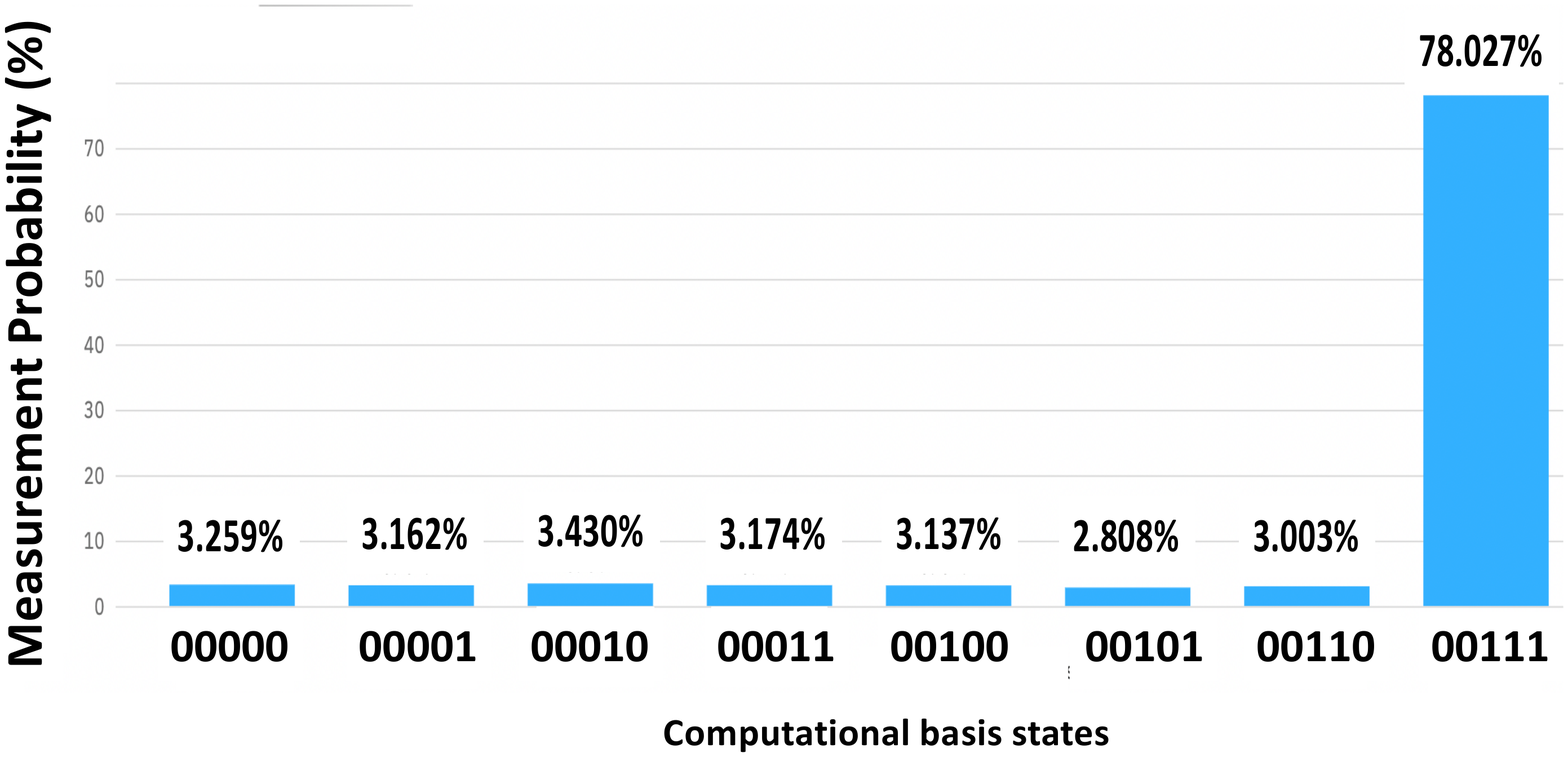}
\vspace*{-10 mm}
\includegraphics[width=0.45\textwidth, height=2.8 in]{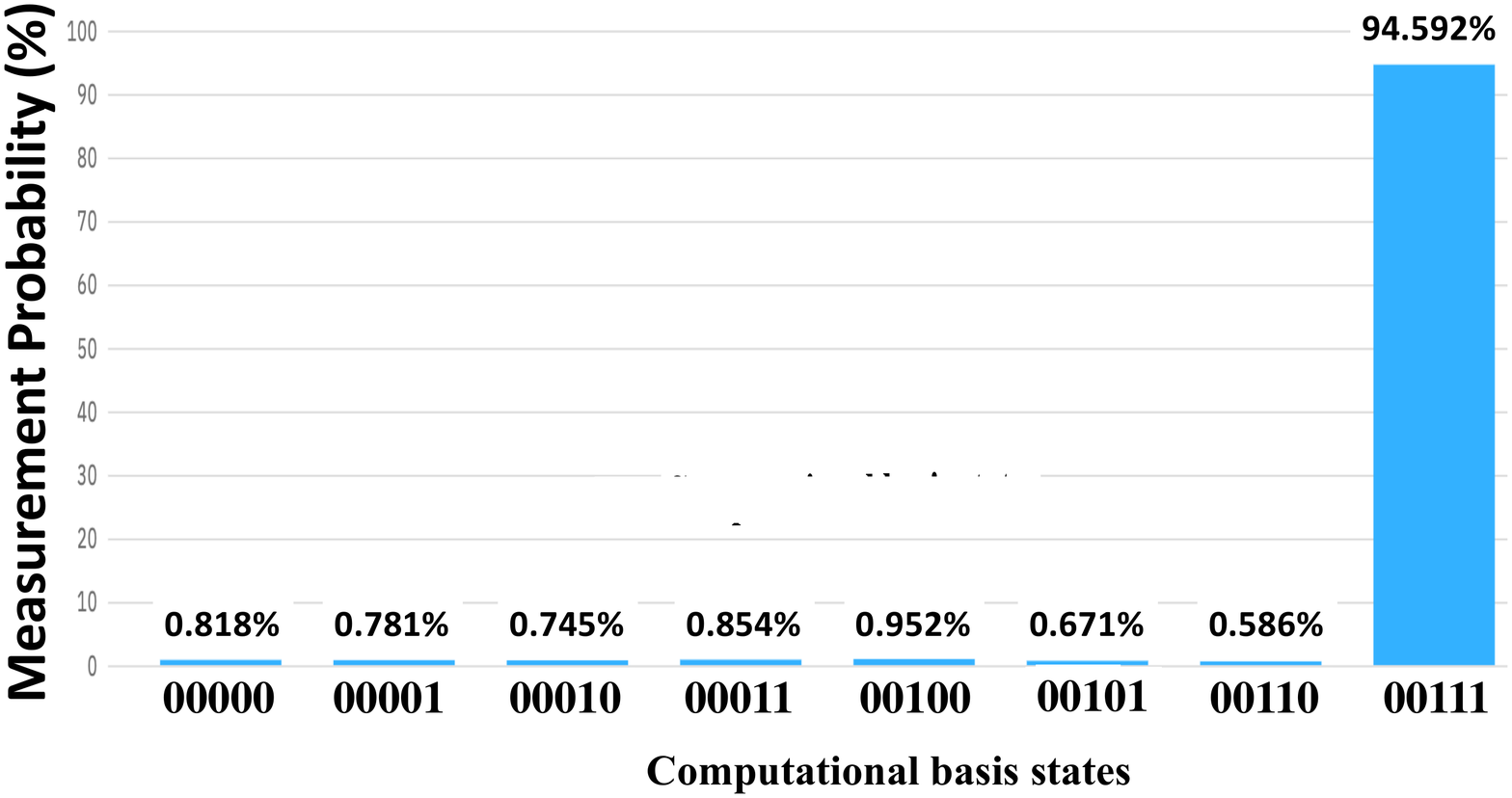}
\includegraphics[width=0.45\textwidth, height=2.8 in]{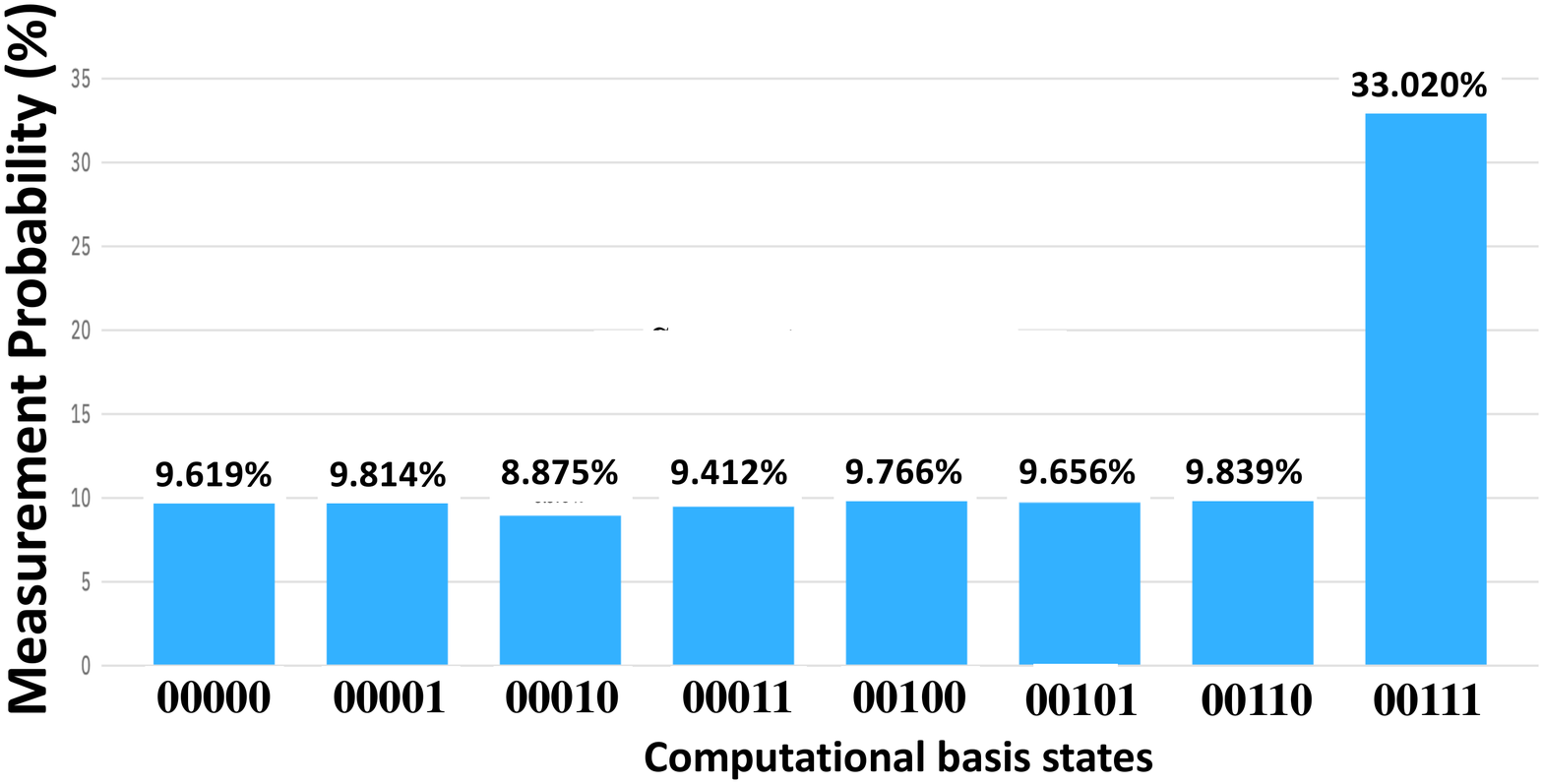}
\caption{{\bf Probability of target state measurement with multiple iterations.} Optimized probability of target state measurement occurs after $\frac{\pi}{4}\sqrt{2^n}$ iterations, rounded to the nearest integer. For a three-qubit quantum system the optimal number of iterations is $\frac{\pi}{4}\sqrt{2^3} \approx 2.22 \approx 2$.}
\label{fig:itr}
\end{figure}

\subsection{Database}
The database contained the transverse momenta (the component of momentum that is perpendicular to 
the beam line of collided particles) of detected leptons. Leptons were recorded with their event number. 
To distinguish between leptons in the same event, "instance" values were assigned. Any first lepton 
detected in its event had an instance = 0; the second detected lepton in its event, an instance = 1; for 
the third, instance = 2, and for the fourth, instance = 3. Figure~\ref{fig:OpenData} is a sample from the 
database and displays the arrangement of the data. Thus, the presence of a lepton instance value of 
three in the database indicated that four leptons were detected following one collision.

To apply the quantum search algorithm to the ATLAS data, the data was encoded into the quantum 
circuit. Classical data, in groups of four entries, was placed into the quantum states of a five-qubit 
quantum system as follows. As each lepton received an "instance" value for the event in which it was 
detected by ATLAS, the marked state corresponded to instance = 3. The search was run on the five-qubit 
quantum system, with q[0] and q[1] defining the index of the lepton in its group of four leptons searched 
by GA, and with q[2] and q[3] keying in the value of the lepton "instance" with respect to event, as in the 
database. Qubit q[4] served as an ancillary qubit. A Jupyter Notebook was used to compose code to run 
on IBM devices. An interactive simulator called Quirk was employed for developmental purposes. Quirk allows developers to drag quantum gates onto a circuit and 
demonstrates the direct effect on the amplitude, probability, and representation of each state at the end 
of the circuit. Quirk provides real-time simulated results as the circuit changes, rather than requiring 
users to run the circuit and a final measurement step in order to obtain the results of the simulation.

\begin{widetext}
\begin{center}	
\begin{figure}[h!] 
\includegraphics[width=0.55\textwidth, height = 4. in]{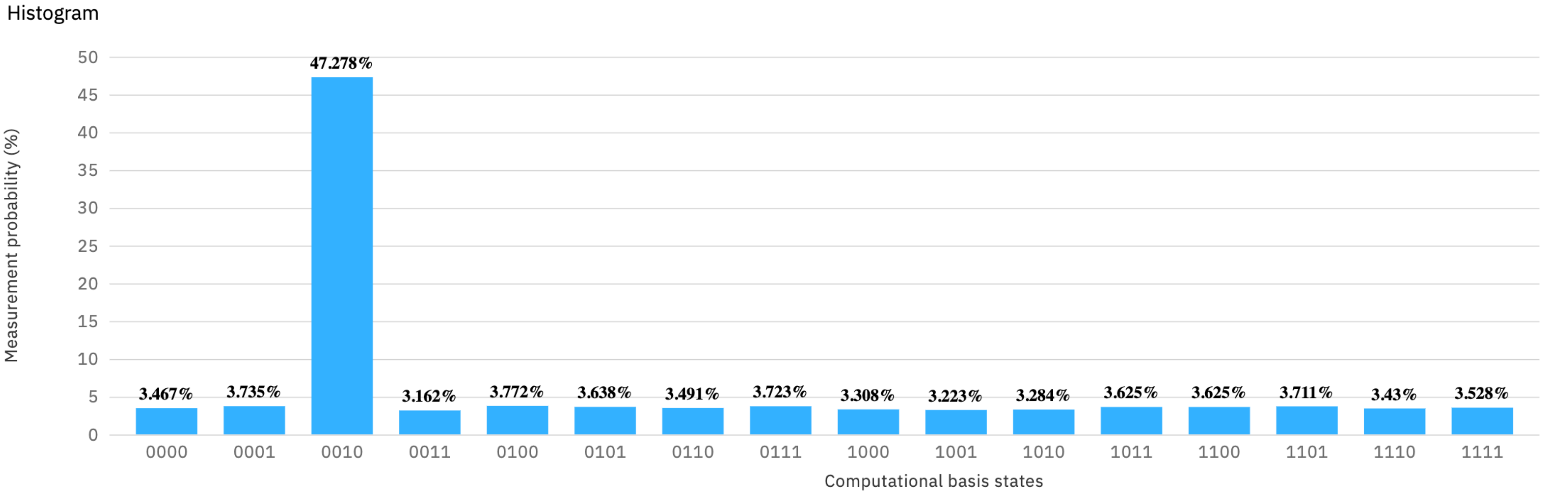}
\includegraphics[width=0.45\textwidth, height=2. in]{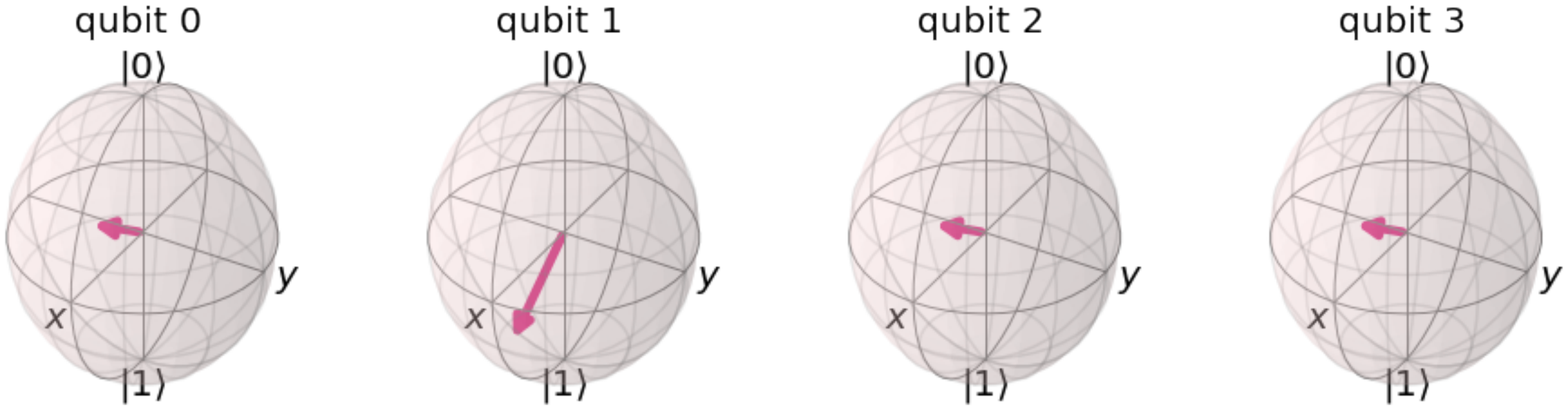}
\vspace*{-6 mm}
\caption{{\bf Four-qubit quantum search and qubit representation.} The algorithm yielded a distinct peak in probability of target state measurement (left) when run on four qubits. The final multi-qubit state that is conducive for a successful search is shown in Bloch Sphere representations of each qubit (right)}
\label{fig:fourqubit}
\end{figure}
\vspace*{-4 mm}
\end{center}	
 \end{widetext}

Our code was run on each group of data entries thousands of times to yield accurate probability 
distributions for the final state of the quantum system. All iterations on the data subsets not containing 
the marked lepton state yielded a quantum system in equal superposition and thus an equal probability 
of measurement in each state. Those iterations on groups of four data entries containing an index value 
of 3 were hypothesized to leave the quantum system with a higher probability---mathematically, 100\%---of 
measuring the marked state with index = 3. In order to solely yield the data entry (row) number in the 
database and the transverse momentum of exclusively leptons with index = 3, the code yielded these 
values for those states with a probability of being measured $\ge$ (.80). The algorithm was first run and 
tested on IBM's ibmq\_qasm\_simulator backend before it was run on several IBM quantum computers.

\begin{figure}[h!] 
\centering
 \vspace*{-1.0 mm}
 \includegraphics[width=0.4\textwidth]{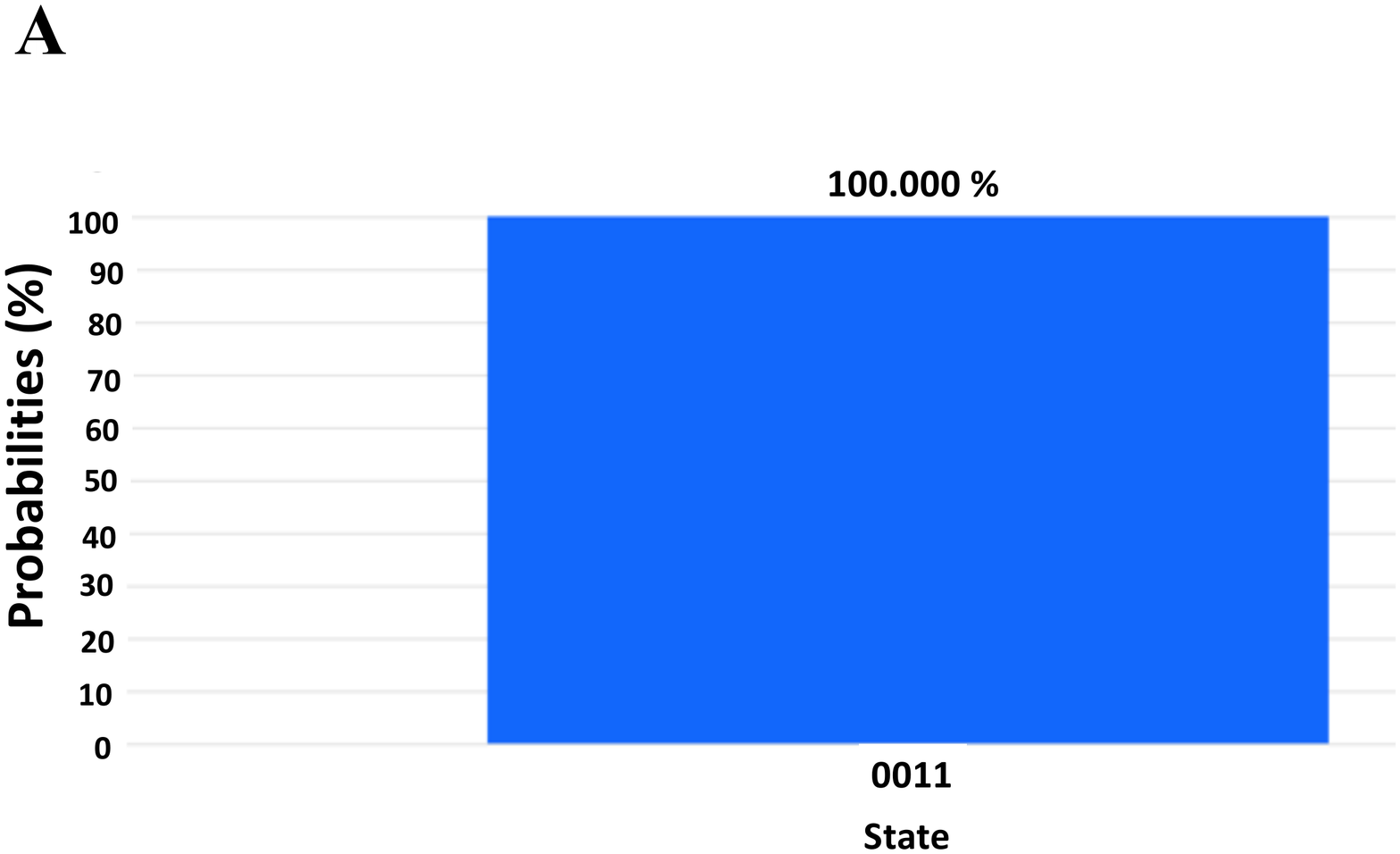}
 \vspace*{-10 mm}
  \includegraphics[width=0.4\textwidth]{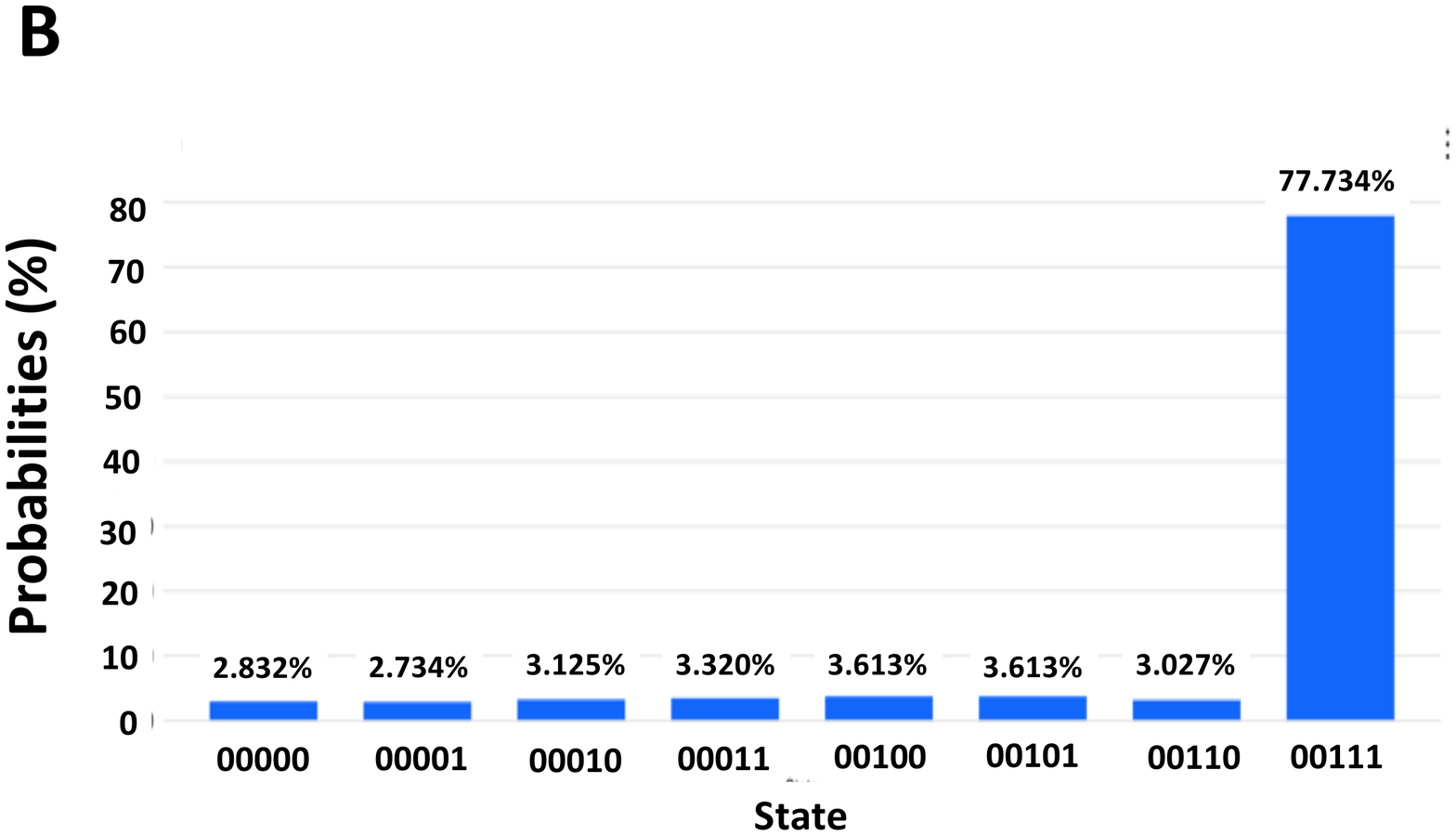}
   \vspace*{-10 mm}
 \includegraphics[width=0.4\textwidth]{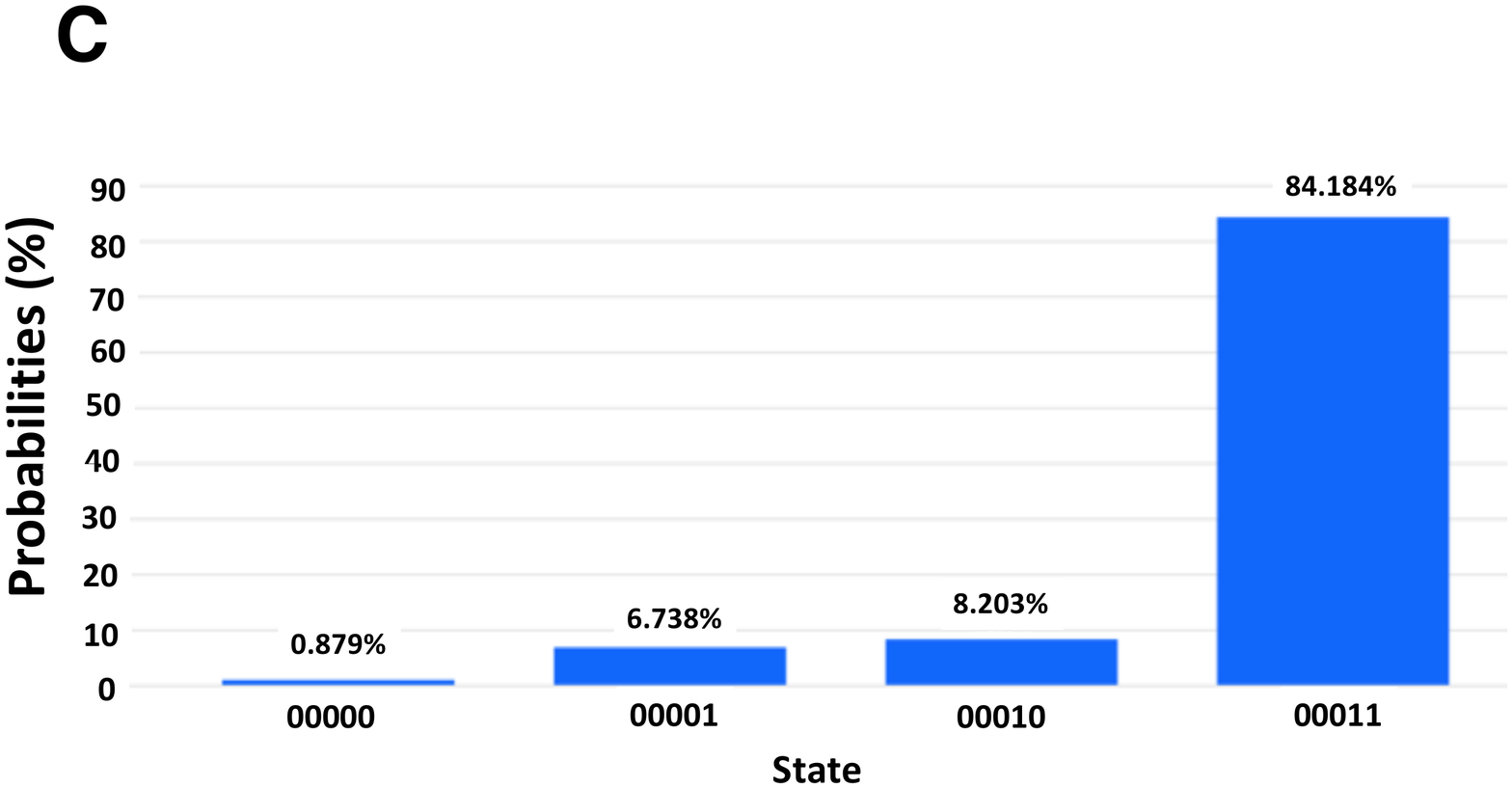}
  \vspace*{-15 mm}
 \includegraphics[width=0.4\textwidth]{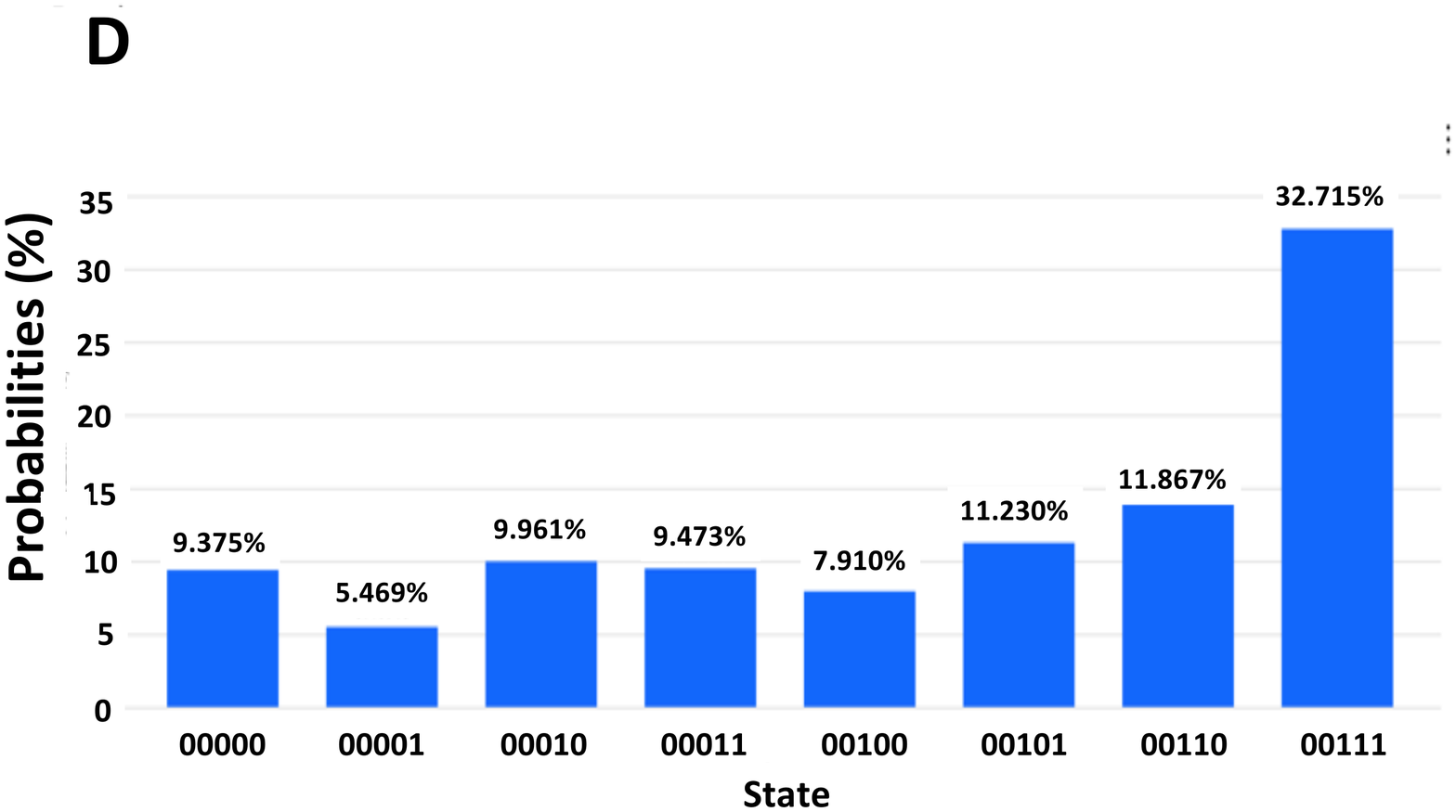}
\vspace*{12 mm}
\caption{{\bf Grover’s algorithm applied to two- and three-qubit systems in Qiskit.}
The probability of each two-qubit (A, C) and three-qubit (B, D) state is presented upon measurement after one iteration of GA on the Qiskit simulator (A, B) and on real IBM QCs (C, D).}
 \label{fig:GA23}
\end{figure}

\section{Results}
\subsection{Grover’s Algorithm Execution}
The probability distribution of each quantum state's measurement following steps and iterations of 
Grover's algorithm reflects the manipulation of amplitude to yield a peak in probability of target state 
measurement. A three-qubit quantum chip placed in an equal superposition demonstrates a 1/8 probability 
of each possible state being measured. After one iteration of GA, there is a probability of 0.78 that the 
chip is measured in the target state (Fig~\ref{fig:itr}). Following two iterations of Grover's algorithm on 
three qubits, the target state is selected with a probability of 0.94 on the QC simulator. Our results 
    shown in Fig~\ref{fig:itr} corroborated models for the optimization of GA iterations, which indicate that the 
ideal number of iterations on n qubits is $\frac{\pi}{4}\sqrt{2^n}$ assuming no hardware error 
\cite{tightbounds}. When three iterations of GA were run on a three-qubit 
system, the correct state was selected with a probability of .33. Fig~\ref{fig:itr} shows the results of 
Grover's algorithm with a varying number of iterations on three qubits. The optimization of target state 
measurement probability is a key element of conducting quantum searches with Grover's algorithm; this 
factor was considered along with quantum decoherence when applying the quantum search to ATLAS 
data.

Grover's algorithm was run on a varying number of qubits. The algorithm continued to yield a distinct 
peak in target probability after one iteration, as seen in Fig~\ref{fig:fourqubit}. Yet, as the number of 
qubits and thus possible states increased, correct measurement was made with a lesser probability. 
Fig~\ref{fig:fourqubit}A shows a histogram of measurements made after one iteration of GA on four 
qubits. As shown, one iteration of GA leaves the four-qubit system with a .47 probability of correct 
measurement, while one iteration on three qubits results in measurement of the target state with a 
probability of .78, as shown in Fig~\ref{fig:itr}. As $\frac{\pi}{4}\sqrt{2^n}$ GA iterations optimize 
selection accuracy, an increasing number of iterations is needed to sufficiently raise target state 
probability as the number of qubits increases. However, when run on real devices, longer quantum 
codes fall prey to decoherence, and iteration number is thus best kept to a minimum for a more 
effective quantum search. 

\begin{figure}[t] 
\centering
\includegraphics[width=0.45\textwidth]{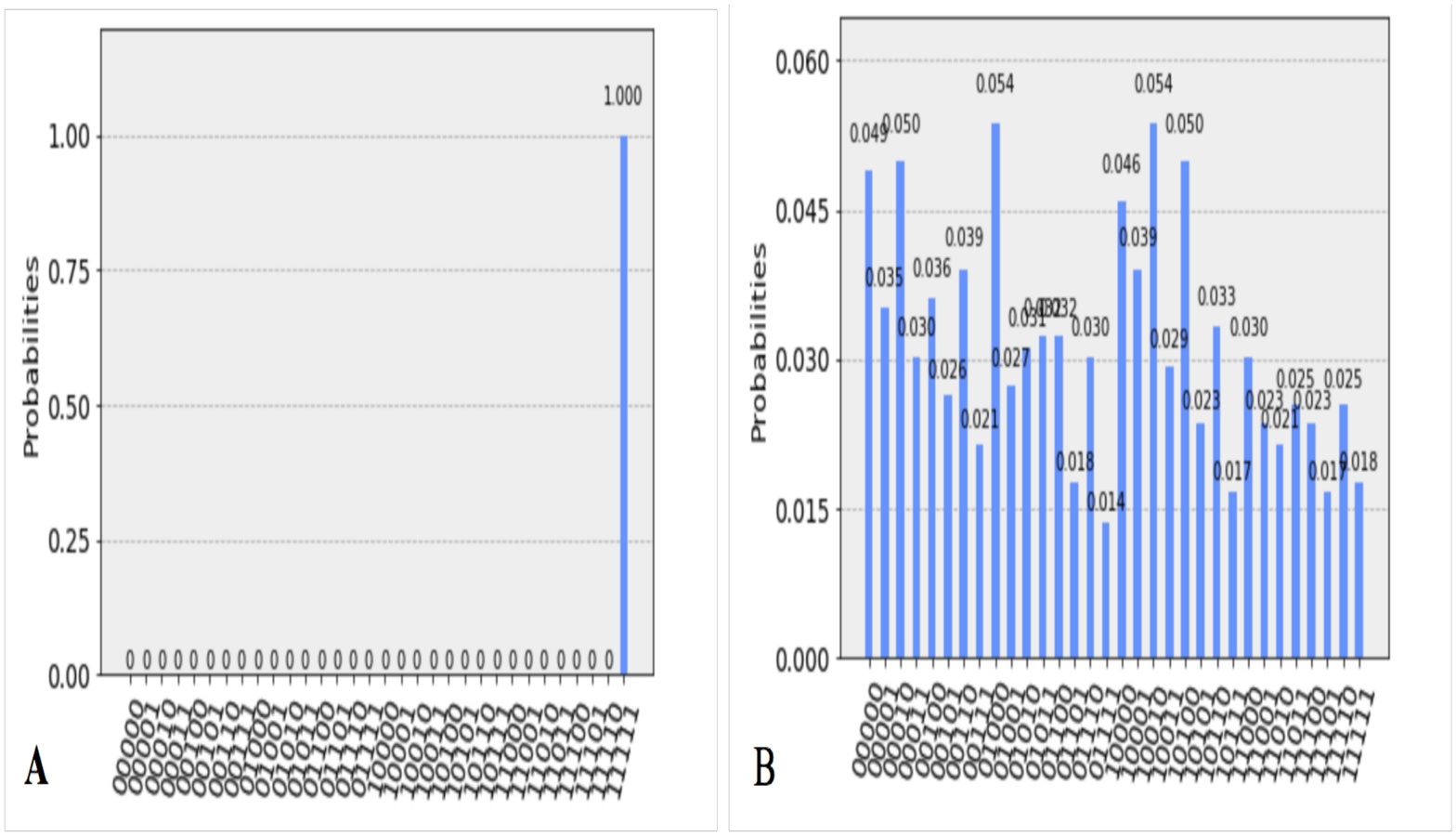}
\caption{{\bf Histogram of probabilities by quantum state with the ATLAS data encoded.}  (Left): The group of four data entries included the target lepton index = 3. The algorithm was run on Qiskit's ibmq\_qasm\_simulator backend, a classical simulation. B (Right): This group of four data entries contained the target lepton index = 3. The effects of decoherence can be seen. The algorithm was run on the quantum device ibmq\_16\_melbourne. }
\label{fig:Histprob}
\end{figure}

Grover's algorithm was executed on increasing numbers of qubits on real IBM quantum computers, 
which demonstrated a similar peak distinction once measured. The probability distributions of runs on 
QCs were compared to those of simulated runs. As seen in Fig~\ref{fig:GA23}, 
the peak in probability is slightly less distinct after runs on real QCs due to quantum decoherence. 

\subsection{Grover's Algorithm Application to LHC Data: First Encoding}
\begin{figure}[bp!] 
\centering
\includegraphics[width=0.48\textwidth, height = 1.5 in]{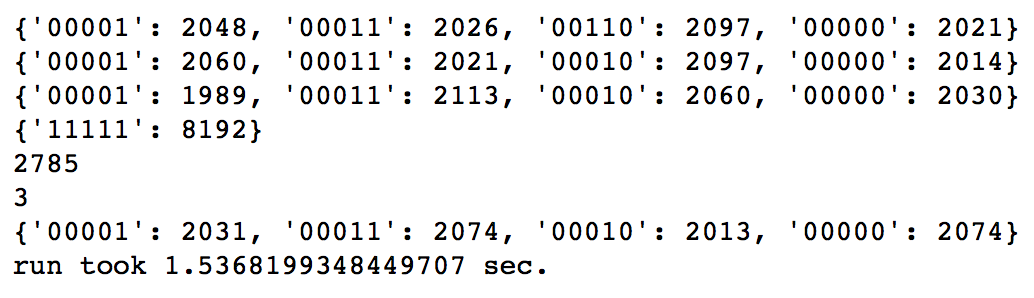}
\caption{{\bf Final state measurement counts for five groups of four data entries, each searched by GA.} The fourth group contains the target lepton index = 3, and its corresponding state was measured all 8192 times on the simulator backend.}
\label{fig:FSM}
\end{figure}

\begin{figure}[t!] 
\centering
 \includegraphics[angle=-90, origin=c,width=0.5\textwidth]{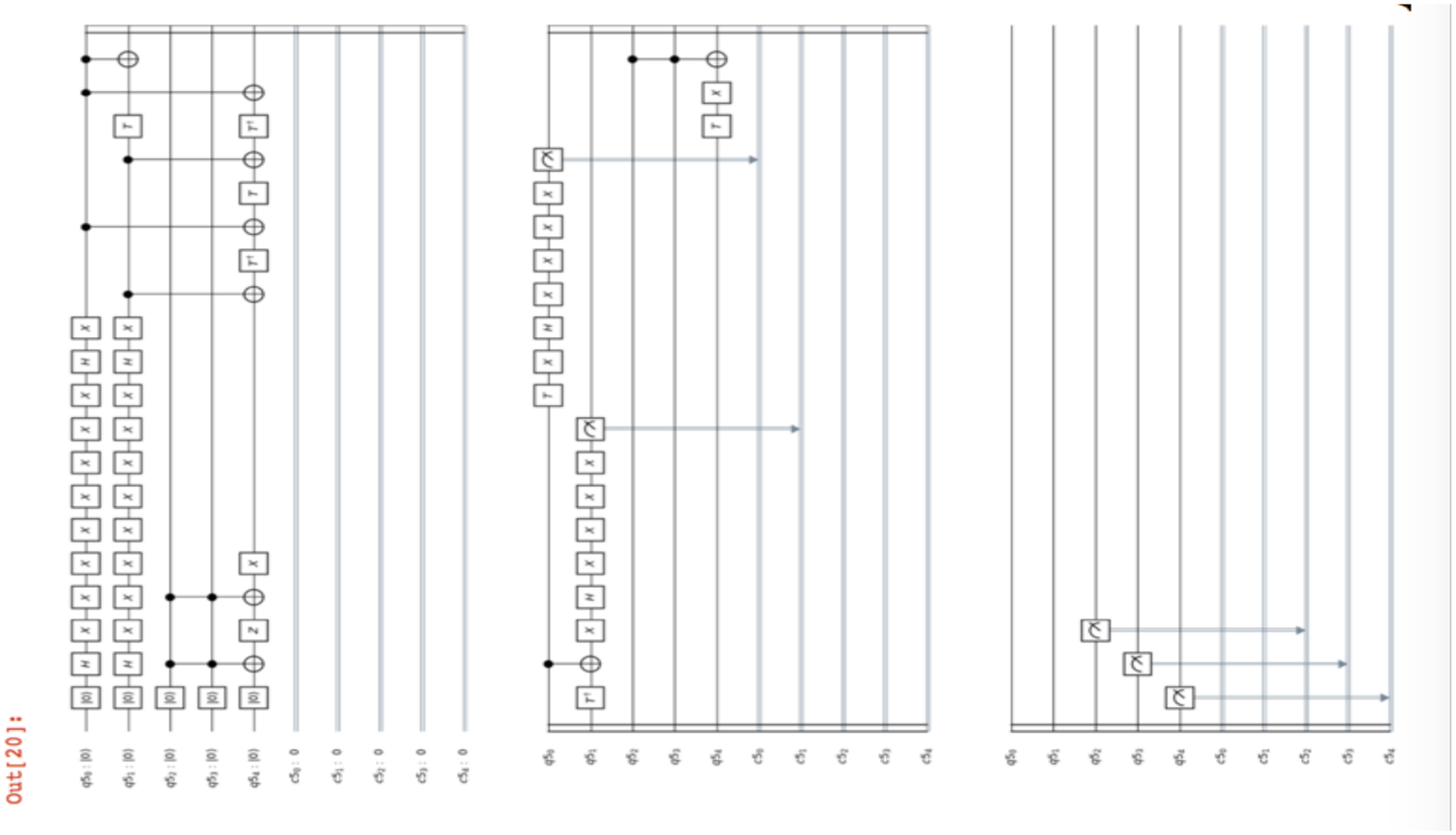}
\caption{{\bf  Quantum circuit.}  A quantum circuit generated in the selection process.}
\label{fig:QC}
\end{figure}

The classical simulator backend ibmq\_qasm\_simulator yielded the target state with a probability 
of 1.0 (100\%), as seen in Fig~\ref{fig:Histprob} and Fig~\ref{fig:FSM}. The search run on the real 15-qubit quantum computer ibmq\_16\_melbourne demonstrated a limited peak amplification, the consequence of quantum decoherence (Fig~\ref{fig:Histprob}).
The code that was written generates distinct quantum circuits depending upon the values in the database. One quantum circuit, a map of all the quantum gates implemented in one run of Grover's algorithm on the ATLAS data, is shown in Fig~\ref{fig:QC}.

\section{Grover's Algorithm Application to LHC Data: Second Procedure}
After decreasing the number of quantum gates and running the code on several devices with the lowest error rates, we concluded that with the current level of QC engineering, decoherence to the degree of limiting meaningful results occurred with the amount of quantum gates our primary method of encoding the data requires. A different method of encoding data into the quantum computer was developed for the purpose of obtaining successful results with the latest existing quantum computers rather than anticipating advancements in their engineering. We sought a method successful beyond theoretical quantum information. Our second method of encoding classical data to be searched requires fewer 
quantum gates and saw successful results on both the simulator and real quantum computers. 

The casting of unsorted data---in groups of 8---into the quantum computer began by defining qubits q[0] and q[1] as the value register. These two qubits encode the instance values of the database, which, 
as previously, are in the discreet set \{0, 3\}, as did q[2] and q[3] in the primary encoding method already 
described. In this second method, q[2], q[3], and q[4] served to facilitate a binary encoding of index 
values contained in the discreet set \{0, 7\}. The procedure is as follows:

If an instance value of 3 is contained in the set, the state $\ket{\_\_\_11}$ is marked, where "\_\_\_" denotes the unspecified states of q[4], q[3], and q[2] in that order and "11" indicates q[1] and q[0] are both in the $\ket{1}$ state.  In the following three qubits q[2], q[3], and q[4], X gates are placed to create a binary encoding. 
A function keys in the index of the data, in \{0, 7\}, on q[2], q[3], and q[4] using a binary encoding. All qubits are prepared in the $\ket{0}$ state, and the single-qubit X gate transforms the state of a qubit from $\ket{0}$ to $\ket{1}$ and vice-versa. For the index value of 0, no X gates are placed; for the index value of 1, an X gate is placed on q[2] alone, leaving the q[4]-q[3]-q[2]... state to be $\ket{001...}$; for the index value of 2, an X gate is placed on q[3] alone, placing the three index value encoded qubits in the $\ket{010...}$ state; for the index value of 3, an X gate is placed on q[4] alone, and so on. This "Index" function contains a subfunction f where, if the database value is 0, 1, or 2, then f(x)=Index$^{-1}$, but if the database value is 3, then f(x) is an identity transformation. The result is a series of three 0s and 1s, representing the $\ket{0}$ and $\ket{1}$ states of qubits, that corresponds to the specific index value, in binary form, where the target instance value of 3 is found. The final measured state can thus be read as $\ket{vwxyz}$ where $\ket{vwx}$ is the measured three-qubit state of q[4], q[3], and q[2] in that order, and $\ket{yz}$ is the measured two-qubit state of q[1] and q[0]. Both $vwx$ and $yz$ are series of 0's and 1's, where $yz$ represents the searched database value in binary, and $vwx$ represents the corresponding index value in binary. Thus, the selected state indicates both the searched instance value and the index value where the target instance value is found. 

\begin{figure}[tp!] 
\vspace*{-10 mm}
\centering
\includegraphics[width=0.5\textwidth]{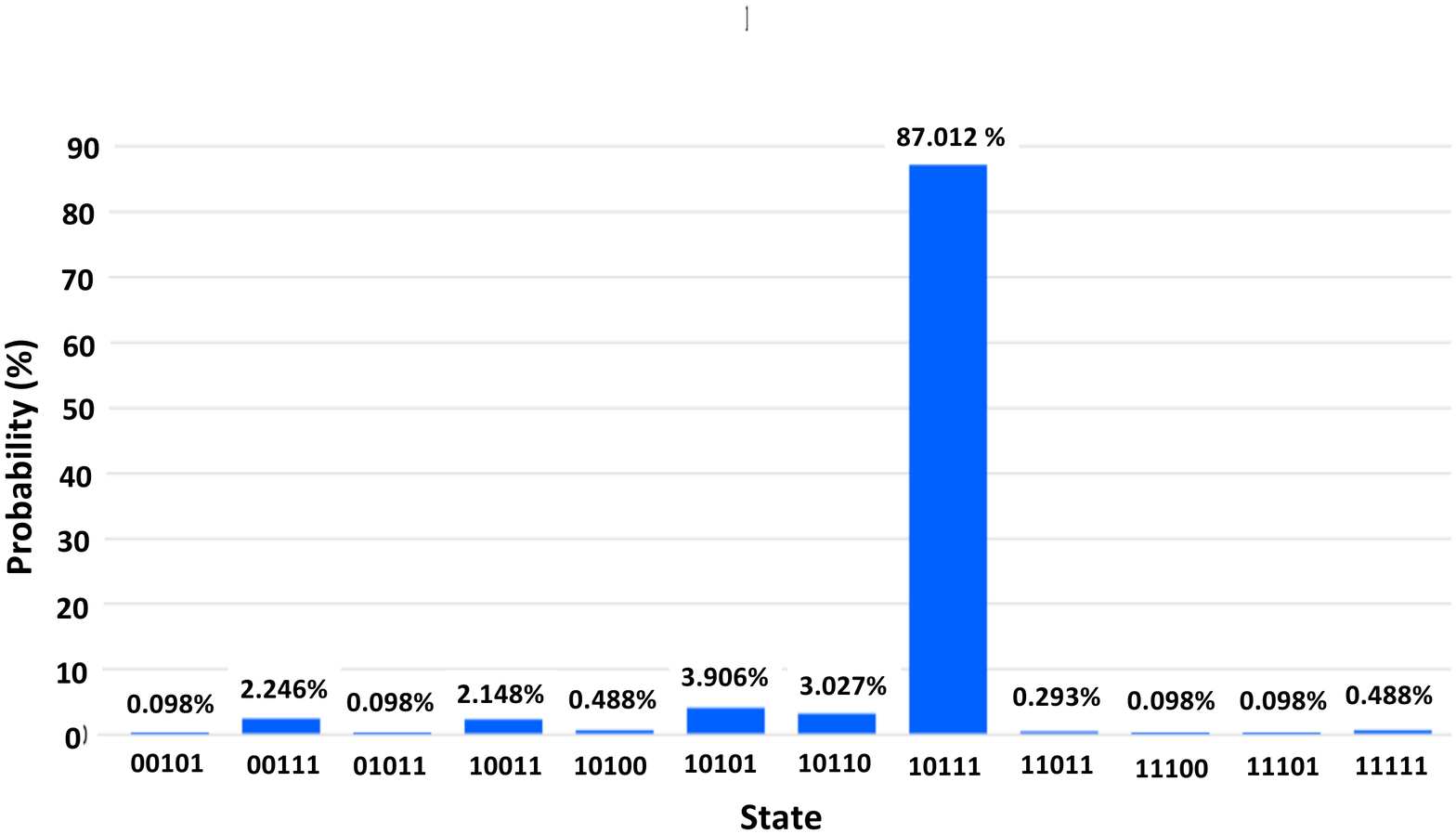}
\vspace*{-10 mm}
\caption{{\bf Histogram of final state probabilities after 8,192 runs on the quantum computer ibmq\_vigo.}  The correct state was selected with a probability of 87.012 \%, or 7,128 of 8,192 runs. The 20 of 32 states not shown in the histogram were states in which the system was not measured on any run.}
\label{fig:Encoding-2}
\end{figure}

\section{Results:  Second Procedure}
The search using the second encoding method was run on QCs and QC simulators, all obtaining results 
that reflect success. As with the first encoding method, the simulator selected the correct state with a 
probability of 1.0, or 100\%. On the quantum computer ibmq\_vigo, the five-qubit system was measured 
in the correct state 87.012\% of the time, as seen in Fig~\ref{fig:Encoding-2}. With limited quantum 
decoherence owing to the modified, more efficient encoding method, the runs of the real quantum 
computer were successful.

\section{Discussion}
The application of a quantum search to data from $pp$ collisions at 13 TeV collision energy is novel, 
and the 100\% correct selection rate on the simulator demonstrates that, mathematically, the 
maneuvers on qubit amplitudes successfully yielded the correct result upon measurement. Yet at first, when 
the primary encoding method was run on the 15-qubit backend ibmq\_16\_melbourne, where the 
quantum speed-up is reaped, the peak amplification was reduced as compared to the simulator 
devices. 

The difference in amplification is explained by quantum decoherence---the interaction of qubits with their 
environment that randomizes results in quantum computing---reducing the peak as seen in 
Fig~\ref{fig:Histprob}. Single qubit (U2) and inter-qubit (CNOT) error rates are shown in 
Fig~\ref{fig:Errors}. While relatively low for each gate, the error rates compound over the quantum 
circuit. The more gates used in a quantum code, the greater the compound error rate becomes, and the 
greater the likelihood that qubits interact adversely with their environment. Current research efforts are 
directed toward combating decoherence. The improving engineering of quantum computers offers 
promise in limiting error in the execution of longer quantum codes. 

A second encoding method was developed to decrease the number of gates and thus the compounded 
error rate of the quantum search. The resulting decrease in discrepancy between the results of the QC 
simulator and those of the real quantum devices is a reflection of mitigated error. In a simulated 
environment, without quantum decoherence, GA successfully and effectively searched ATLAS Open 
Data for events with four leptons---with 100\% accuracy. GA selected the target state with high probability 
when run on a real QC, demonstrating the viability of this extremely advantageous search method. The 
presented results are promising in databases directly formatted to be searched by quantum computers; with 
LHC databases constructed in this way, quantum computers will expedite the data sorting, search, and 
analysis processes.

\section{Conclusion}
Particle production events in the ATLAS detector at CERN's LHC are the result of energy deposits in 
the calorimeter detector cells and charged particle hits in the tracking detectors.  Combinations of 
algorithms developed in the collaboration are used to make the charged particle identifications, which 
are of muons and electrons in the study described here. The resulting events from this process are 
analyzed using mainly classical algorithms run on classical computers.  

\begin{figure}[h!] 
\centering
\includegraphics[width=0.45\textwidth]{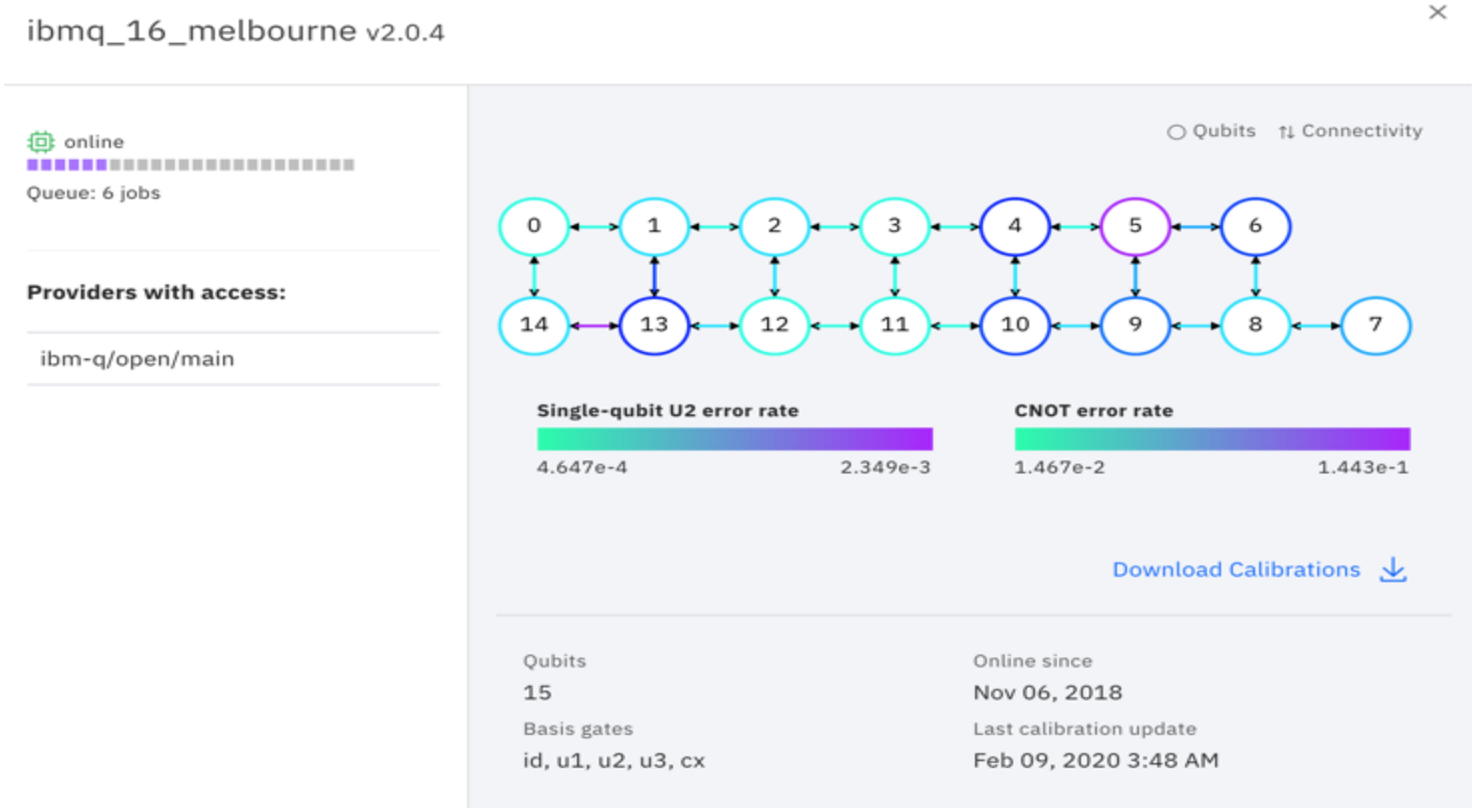}
\caption{{\bf A diagram of single- and multi-qubit error rates in ibmq\_16\_melbourne, provided by IBMQ.}}
\label{fig:Errors}
\end{figure}

The HL-LHC will produce increased event rates and a more challenging environment for selection of rare signal events of interest to scientists against a large bevy of scientifically inconsequential background.  Selection of rare, interesting events as described in this article requires algorithms that can select and then format the detector output into a form that can be recognized and used by quantum algorithms, such as Grover's search algorithm.  We have demonstrated one method as an example---GA applied to properly formatted Higgs boson events in the ATLAS collaboration.  The data was converted from formats conventionally used in classical analysis to quantum formats that are used by a quantum algorithm.  GA applied to the small database of events described here demonstrates a way forward for use with varied databases composed of unsorted particle physics data events. 

The 100\% success rate of Grover’s algorithm in the simulated environment in searching the ATLAS Open Data for four leptons in the same event substantiates the theoretical viability of GA. The algorithm's success on quantum computing simulators extends to the runs on real quantum computers, where the benefits of quantum computing to efficacy lie. Our findings indicate that the use of quantum computing in particle physics would contribute to the acceleration of scientific discovery by, as shown via both the simulation as well as quantum computers, improving data selection at the LHC. There was mild discrepancy between the results on the quantum computers and those on the simulators---a product of quantum decoherence; see Fig~\ref{fig:Errors}. With a strategically planned encoding process and with the reduction of quantum gates in our code, decoherence was mitigated. The result is a valuable set of results that offer promise to both quantum computing and particle physics, and to the application of the latter to the former.

\section{Acknowledgements}
AEA would like to thank Theodota Lagouri and Andreas Pfeiffer for their invaluable assistance and 
support. OKB gratefully acknowledges funding support from the Department of Energy QuantISED 
Award DE-SC0019592.

\bibliographystyle{plain}

\begin{thebibliography}{9}

\bibitem{Grover_1996}
K. Grover, "A fast quantum mechanical algorithm for database search". Proc. 28th Annual
ACM Symposium on the Theory of Computing (1996); arXiv:quant-ph/9605043 (1996).

\bibitem{Korepin_2005}
V.E. Korepin and L.K. Grover, "Simple algorithm for partial quantum search", Quant. Info. Proc.
{\bf 5}, 5 (2006); arXiv:quant-ph/0504157 (2005).

\bibitem{atlas2012observation}
G. Aad et.al., "Observation of a new particle in the search for 
the Standard Model Higgs boson with the ATLAS detector at the LHC",
Phys. Lett. B716, 1 (2012).

\bibitem{cms2012observation}
S.Chatrchyan et.al., "Observation of a new boson at a mass of 125 GeV 
with the CMS experiment at the LHC", Phys. Rev. B716, 30 (2012).

\bibitem{Marciano:2013t1}
H. Davoudiasl et.al., "Higgs decays as a window into the dark sector"
Phys. Rev. D88, 015022 (2013).

\bibitem{BNL2012dark}
H. Davoudiasl, H.S. Lee, and W.J. Marciano, "Dark Z implications for parity
violation, rare meson decays, and Higgs physics", Phys. Rev. D85. 115019 (2012).

\bibitem{Curtin:2013fra}
D. Curtin et.al., "Exotic Decays of the 125 GeV Higgs Boson", 
Phys. Rev. D90, 075004 (2014).

\bibitem{gopalakrishna2008higgs}
S. Gopalakrishna and S. Jung and J. D. Wells, "Higgs boson decays to four fermions through an
abelian hidden sector", Phys. Rev. D78, 055002 (2008).

\bibitem{zdark:ATLAS}
G. Aad et.al., "Search for new light gauge bosons in Higgs boson decays to four-lepton final 
states in pp collisions at $\sqrt s = 8$ TeV with the ATLAS detector at the LHC",  Phys. Rev. 
D92, 092001 (2015).

 \bibitem{2011:Strubell}
 E. Strubell, "An introduction to quantum algorithms.", University of Massachusetts,
 Lecture Notes (2011).

 \bibitem{2002:Diao}
 Z. Diao, M.S. Zubairy, G. Chen, "A quantum circuit design for grover’s algorithm.", 
 Z.  Naturforschung a 57(8), 701–708 (2002).
 
 \bibitem{tightbounds}
M. Boyer, G. Brassard, P. Hoyer and A. Tapp, "Tight bounds on quantum searching", Fortschr. Phys., 46, pp. 493-506,1998.
 
\end{thebibliography}

\end{document}